\documentclass[showpacs,preprintnumbers, nofootinbib]{revtex4}
\usepackage{fontenc}
\usepackage{graphicx}
\usepackage[dvips]{hyperref}
\usepackage{amsmath}

\begin{document}

\title{Dirac quasinormal modes of Chern-Simons and BTZ black holes with
torsion}
\author{Ram\'{o}n Becar}
\email{rbecar@uct.cl}
\affiliation{Departamento de Ciencias Matem\'{a}ticas y F\'{\i}sicas, Universidad Cat\'{o}%
lica de Temuco, Montt 56, Casilla 15-D, Temuco, Chile}
\author{P. A. Gonz\'{a}lez}
\email{pablo.gonzalez@udp.cl}
\affiliation{Facultad de Ingenier\'{\i}a, Universidad Diego Portales, Avenida Ej\'{e}%
rcito Libertador 441, Casilla 298-V, Santiago, Chile.}
\author{Y. V\'{a}squez.}
\email{yerko.vasquez@ufrontera.cl}
\affiliation{Departamento de Ciencias F\'{\i}sicas, Facultad de Ingenier\'{\i}a, Ciencias
y Administraci\'{o}n, Universidad de La Frontera, Avenida Francisco Salazar
01145, Casilla 54-D, Temuco, Chile}
\affiliation{Departamento de F\'{\i}sica, Facultad de Ciencias, Universidad de La Serena,\\ 
Avenida Cisternas 1200, La Serena, Chile.}
\date{\today }

\begin{abstract}
We study Chern-Simons black holes in $d-$dimensions and we calculate
analytically the quasinormal modes of fermionic perturbations. Also, we
consider as background the five-dimensional Chern-Simons black hole with torsion
and the BTZ black hole with torsion. We have found that the quasinormal modes
depend on the highest power of curvature present in the Chern-Simons theory,
such as occurs for the quasinormal modes of scalar perturbations. We also show that the effect
of the torsion is to modify the real part of the quasinormal frequencies,
which modify the oscillation frequency of the field for the
five-dimensional case. However, for the BTZ black hole with torsion, the effect
is to modify the imaginary part of these frequencies, that is, the relaxation
time for the decay of the black hole perturbation. The imaginary part of the quasinormal frequencies is negative which guaranties the stability of these black holes under fermionic field perturbations. 
\end{abstract}

\maketitle


\section{Introduction}
The quasinormal modes (QNMs) and their quasinormal frequencies (QNFs) are an important property of black holes and have a long history, \cite{Regge:1957td, Zerilli:1971wd, Zerilli:1970se, Kokkotas:1999bd, Nollert:1999ji, Konoplya:2011qq}. It is known that the
presence of event horizons dampens the vibration modes of a matter field
that evolves perturbatively in the exterior region. In this way, the system
is intrinsically dissipative, i.e., there is no temporary symmetry. In general,
the oscillation frequencies are complex therefore, the system is not
Hermitian. Nevertheless, the oscillation frequency of these modes is
independent of the initial conditions and it only depends on the
parameters (mass, charge and angular momentum) and the fundamental constants
(Newton constant and cosmological constant) that describe a black hole just
like the parameters that define the test field. 

The study of the QNMs gives information about the stability of black holes under matter fields that evolves perturbatively in the exterior region of them, without backreacting on the metric, \cite{Abdalla:2010nq, CuadrosMelgar:2011up,Gonzalez:2012de,Gonzalez:2012xc,Becar:2012bj}. Also, the
QNMs determine how fast a thermal state in the boundary theory
will reach thermal equilibrium according to the AdS/CFT correspondence \cite{Maldacena:1997re}, where the relaxation time of a thermal state of the
boundary thermal theory is proportional to the inverse of the imaginary part
of the QNMs of the dual gravity background \cite{Horowitz:1999jd}.

In the context of black hole thermodynamics, the QNMs allow the quantum area spectrum of the black hole horizon to be studied, as well as the mass and the entropy spectrum. In this regard, Bekenstein \cite{Bekenstein:1974jk} was the first to propose the idea that in quantum
gravity the area of black hole horizon is quantized, leading to a discrete
spectrum which is evenly spaced. Then, Hod \cite{Hod:1998vk} conjectured that the asymptotic
QNF is related to the quantized black hole area, by identifying the vibrational frequency
with the real part of the QNFs. However, it is not universal for every
black hole background. Then, Kunstatter \cite{Kunstatter:2002pj} propose that the black hole
spectrum can be obtained by imposing the Bohr-Sommerfeld quantization condition
to an adiabatic invariant quantity involving the energy and the vibrational
frequency. Furthermore, Maggiore \cite{Maggiore:2007nq} argued that
in the large damping limit the identification of the vibrational
frequency with the imaginary part of the QNF could lead to the Bekenstein
universal bound. Then, the consequences of these proposals were studied in several spacetimes.

Commonly the analysis of the QNMs has been carried out on gravitational theories with a Riemannian geometry. However, the incorporation of torsion in the geometry has acquired great interest (for instance, Poincare Gauge Theory of Gravity, Teleparallel Gravity, f(T)-Gravity and f(R,T)-Gravity). In this sense, the simplest gravitational theory that allows spacetime to have torsion is the Einstein-Cartan theory of gravity \cite{cartan, Hehl:1976kj}. Where, the source of torsion should be the spin of matter fields \footnote{Generically, the spin of matter fields is considered to be the source of torsion. However, it can emerges from other sources, see for example \cite{Capozziello:2001mq}.} and the curvature and torsion represent independent degrees of freedom of the gravitational field. In four dimensions, the motion equations give that the torsion is non-propagating, and null in the absence of sources. However, in higher dimensions the Lovelock theories allow a first order formulation \cite{Zumino:1985dp} and the equations of motion admit non null torsion solutions even in the absence of sources, in such a way that it becomes a new propagating degree of freedom \cite{Troncoso:1999pk, Zanelli:2005sa}. Some black holes solutions with torsion in these theories have been studied in \cite{Canfora:2007fw}. In three dimensions, a theory of gravity that includes torsion is known with the name of Mielke-Baekler theory \cite{Mielke:1991nn},
and this model is itself a Chern-Simons theory which includes (along
with the Einstein-Hilbert) the gravitational Chern-Simons term and a translational Chern-Simons term. This model admits as solution a 
 generalization of the BTZ black hole with torsion \cite{Garcia:2003nm}. Three-dimensional gravity with torsion was also considered in Ref. \cite{Giacomini:2006dr}, where the supersymmetric extension in the Chern-Simons formulation was investigated. Also, exact solutions with torsion were analyzed recently in Ref. \cite{Blagojevic:2012ye}. Other interesting
spacetimes with torsion may be found in \cite{Klemm:2007yu, Blagojevic:2008xz, Blagojevic:2008ip,
Blagojevic:2009xw,Vasquez:2009mk}. In regarding the applications to physics it is well known that the introduction of torsion often induces new physical effects and changes the local degrees of freedom of the theory, for example, if the Einstein-Hilbert Lagrangian is the reduction of a higher dimensional model, as the Chern-Simons theories, one can obtain a non null torsion that propagates \cite{Banados:1996yj}.  Moreover, gravity with torsion in 2+1 dimensions has been related with the continuum theory of lattice defects in solid physics \cite{Tresguerres:1992ew, Kawai:1994qw}. On the other hand, the relationship between neutrino oscillation and the torsion was proposed in \cite{Adak:2000tp}. Furthermore, it  was discussed  in \cite{Canfora:2007fw} along with other observables effects of torsion in high energy physics and in solid state physics.

In the context of the AdS/CFT correspondence there are some works, where the authors have incorporated the torsion in the geometry. In this regard, the holographic currents in the five-dimensional Chern-Simons gravity with torsion has been studied in \cite{Banados:2006fe}. The holographic structure of the Mielke-Baekler model has been studied in \cite{Klemm:2007yu} and the holographic aspects of four dimensional gravity with a negative cosmological constant deformed by the Nieh-Yan torsional topological invariant with a spacetime-dependent coeffcient have been examined in \cite{Petkou:2010ve, Leigh:2008tt}. Also, basic aspects of the correspondence have been studied in the framework of 3-dimensional gravity with torsion such as holographic energy-momentum, spin currents and the associated (anomalous) Noether-Ward identities \cite{Blagojevic:2013bu}. 

The particular motivation of this work is to calculate the QNMs for fermionic field perturbations by considering some black holes with torsion, in order to study the stability and the effect of the torsion on Dirac QNMs. Here, we consider as background the BTZ black hole with torsion  \cite{Garcia:2003nm}, and the five-dimensional Chern-Simons black hole with torsion, \cite{Canfora:2007xs}. Also, we consider the $d$-dimensional Chern-Simons black hole without torsion, in order to stablish the effect of the torsion on the Dirac QNMs. Chern-Simons black holes are very interesting static solutions of gravity
theories which asymptotically approach spacetimes of constant negative
curvature (AdS spacetimes). They can be considered as generalizations of the
(2+1)-dimensional black holes in higher-dimensional gravity theories
containing higher powers of curvature. The Chern-Simons black holes of
spherical topology have the same causal structure as the BTZ black holes, and
these solutions have a thermodynamical behavior which is unique among all
possible black holes in competing Lanczos-Lovelock theories with the same
asymptotics. The specific heat of these black holes is positive
therefore, they can always reach thermal equilibrium with their surroundings
and are thus stable against thermal fluctuations. Scalar perturbations, mass and area spectrum and greybody factors of
Chern-Simons black holes have been studied in \cite{Gonzalez:2010vv}. Here,
the authors showed that the rate at which a scalar field in the background
of a Chern-Simons black hole will decay or the rate at which the boundary thermal
theory will reach thermal equilibrium depends on the value of the curvature
parameter $k$. The mass and area spectrum of these black holes have a strong
dependence on the topology of the transverse space and are not evenly
spaced. On the other hand, at a low frequency limit there is a range of modes
which contributes to the absorption cross-section. However, beyond a certain
frequency value, the mass of the black hole does not affect the absorption
cross-section, \cite{Gonzalez:2011du}.

The paper is organized as follows. In Sec. II we give a brief review of the
Chern-Simons theory. In Sec. III we calculate the exact QNMs of fermionic
perturbations with spherical, hyperbolic and plane topology for Chern-Simons
black holes in $d$-dimensions. In Sec. IV we calculate the exact QNMs of
fermionic perturbations for the five-dimensional Chern Simons black hole with
torsion and for the BTZ black hole with torsion. Finally, conclusions
are presented in Sec. V. 

\section{Chern-Simons black holes}
In 1915, David Hilbert proposed an action based on
the metrics and derivatives of this and considered the metric as a unique
fundamental field. The Lagrangian given by the Ricci scalar is invariant
under general transformations of coordinates and it is the only one that
provides second-order equations for the metric. The requirement that the
action be stationary for small variations of the metric gives the field
equations of Einstein (in vacuum). Then, in 1971, D. Lovelock \cite{Lovelock:1971yv}
generalized the problem and found, for an arbitrary dimension, all tensors $%
A_{\mu \nu }$ such as: $A_{\mu \nu }$ is symmetrical, $A_{\mu \nu }$ is a
function of the metric tensor and of its first and second derivatives, and $%
A_{\mu \nu }$ has null divergence. In this formulation, the field equations
for gravitation (in vacuum) are $A_{\mu \nu }$ = 0. It is worth mentioning
that in four dimensions, the Einstein tensor and the metric are the only
solutions, and the field equation is equal to the Einstein equations with
constant cosmology. However, in higher dimensions there are tensors with
the higher power of the Riemann curvature tensor that satisfies the requirements
established, which can be obtained from the Lagrangian of Lanczos-Lovelock,
acknowledging a similar proposition made by C. Lanzcos in 1938 for $d=5$,
with $d$ being the number of spacetime dimensions. Using diferential forms,
specifically the vielbein 1-form and spin connection 1-form, which are related
to the 2-form curvature and torsion, and demanding that the Lagrangian
be: a $d$-form invariant under local Lorentz transformations, a local
polynomial of vielben, the spin connection and its exterior derivative and
built not using the Hodge dual. A. Mardones and J. Zanelli \cite%
{Mardones:1990qc} found that in the absence of torsion the only possible
solution is Lovelock's excepting of Pontryagin's densities, which exist in
even dimensions and are different from zero only if $d=4k$, where $k$ is an
integer. But, being close forms, they can be written locally as total
derivatives, not contributing to the motion equations. The coefficients of
Lovelock's action are arbitrary; however, these can be fixed in such a way that
theories have a unique cosmological constant. The Lanczos-Lovelock (LL)
action \cite{Lovelock:1971yv} in $d-$dimensions can be written as follow 
\begin{equation}
\ I_{k}=\kappa \int \sum_{q=0}^{k}c_{q}^{k}L^{q}~,
\label{dtopologicalaction}
\end{equation}
with 
\begin{equation}
\ L^{q}=\epsilon
_{a_{1}...a_{d}}R^{a_{1}a_{2}}...R^{a_{2q-1}a_{2q}}e^{a_{2q+1}}...e^{a_{d}}~,
\label{polinomio}
\end{equation}
where $e^{a}$ and $R^{ab}$ stand for the vielbein and the curvature two-form
respectively, and $\kappa $ and $l$ are related to the gravitational constant 
$G_{k}$ and the cosmological constant $\Lambda $ through 
\begin{equation}
\ \kappa =\frac{1}{2(d-2)!\Omega _{d-2}G_{k}}~,  \label{definitionk}
\end{equation}
\begin{equation}
\ \Lambda =-\frac{(d-1)(d-2)}{2l^{2}}~,  \label{lambda}
\end{equation}
and $\alpha _{q}:=c_{q}^{k}$ where $c_{q}^{k}=\frac{l^{2(q-k)}}{d-2q}
(_{q}^{k})$ for $q\leq k$ and vanishes for $q>k$, with $1\leq k\leq \lbrack 
\frac{d-1}{2}]$ ($[x]$ denotes the integer part of $x$) and $\Omega _{d-2}$
corresponds to the volume of a unit $(d-2)$-dimensional sphere. Static black
hole-like geometries with spherical topology were found in~\cite
{Crisostomo:2000bb} possessing topologically nontrivial AdS asymptotics.
These theories and their corresponding solutions were classified by a $k$ integer, which corresponds to the highest power of curvature in the
Lagrangian. If $d-2k=1$, the solutions are known as Chern-Simons black holes
(for a review on the Chern-Simons theories see \cite{Zanelli:2005sa}). These
solutions were further generalized to other topologies~\cite{Aros:2000ij}
and can be described in general by a non-trivial transverse spatial
section $\sum_{\gamma }$ of $(d-2)$-dimensions labelled by the constant $
\gamma =+1,-1,0$ that represents the curvature of the transverse section,
corresponding to a spherical, hyperbolic or plane section, respectively. The
solution describing a black hole in a free torsion theory can be written as~
\cite{Aros:2000ij} 
\begin{equation}
\ ds^{2}=-\Big{(}\gamma +\frac{r^{2}}{l^{2}}-\alpha \Big{(}\frac{
2G_{k}\sigma }{r^{d-2k-1}}\Big{)}^{\frac{1}{k}}\Big{)}dt^{2}+\frac{dr^{2}}{
\Big{(}\gamma +\frac{r^{2}}{l^{2}}-\alpha \Big{(}\frac{2G_{k}\sigma }{
r^{d-2k-1}}\Big{)}^{\frac{1}{k}}\Big{)}}+r^{2}d\sigma _{\gamma }^{2}~,
\label{dtopologicalmetric}
\end{equation}
where $\alpha =(\pm 1)^{k+1}$ and the constant $\sigma $ is related to the
horizon $r_{+}$ through 
\begin{equation}
\ \sigma =\frac{r_{+}^{d-2k-1}}{2G_{k}}(\gamma +\frac{r_{+}^{2}}{l^{2}}
)^{k}~,  \label{relationr}
\end{equation}
and to the mass $M$ by 
\begin{equation}
\ \sigma =\frac{\Omega _{d-2}}{\Sigma _{d-2}}M+\frac{1}{2G_{k}}\delta
_{d-2k,\gamma }~,  \label{relationM}
\end{equation}
where $\Sigma _{d-2}$ denotes the volume of the transverse space. As can be
seen in (\ref{dtopologicalmetric}), if $d-2k\neq 1$ the $k$ root makes the
curvature singularity milder than the corresponding black hole of the same
mass. At the exact Chern-Simons limit $d-2k=1$, the solution has a 
structure similar to that of like the (2+1)-dimensional BTZ black hole with a string-like
singularity.

\section{Fermionic quasinormal modes of d-dimensional Chern-Simons black holes}

Quasinormal modes of fermionic perturbations are governed by the Dirac
equation. We will determine the quasinormal modes by imposing modes ingoing
at the horizon, since nothing can escape from the black hole (classically)
and, as the black hole is asymptotically AdS, we will consider that the flux
of the field vanishes at infinity. The metric of Chern-Simons theories is 
\begin{equation}
\ ds^{2}=-f(r)^{2}dt^{2}+\frac{1}{f(r)^{2}}dr^{2}+r^{2}d\sigma _{\gamma
}^{2}~,  \label{metric2}
\end{equation}
with 
\begin{equation}
f(r)^{2}=\frac{r^{2}}{l^{2}}-\mu ~,
\end{equation}
where we have defined 
\begin{equation}
\mu =\alpha (2\sigma G_{k})^{\frac{1}{k}}-\gamma~,
\end{equation}
and the horizon is located at 
\begin{equation}
\ r_{+}=l\sqrt{\mu }~.  \label{horizon}
\end{equation}

A minimally coupled fermionic field to curvature in the background of the
Chern-Simons black hole in $d-$dimensions is given by the Dirac equation 
\begin{equation}
\left( \gamma ^{\mu }\nabla _{\mu }+m\right) \psi =0~,
\end{equation}
where the covariant derivative is defined as 
\begin{equation}
\nabla _{\mu }=\partial _{\mu }+\frac{1}{2}\omega _{\text{ \ \ \ }\mu
}^{ab}J_{ab}~,
\end{equation}
and the generators of the Lorentz group $J_{ab}$ are 
\begin{equation}
J_{ab}=\frac{1}{4}\left[ \gamma _{a},\gamma _{b}\right] ~.
\end{equation}
The gamma matrices in curved spacetime $\gamma ^{\mu }$ are defined by 
\begin{equation}
\gamma ^{\mu }=e_{\text{ \ }a}^{\mu }\gamma ^{a}~,
\end{equation}
where $\gamma ^{a}$ are the gamma matrices in flat spacetime. In order to
solve the Dirac equation we use the diagonal vielbein 
\begin{equation}
e^{0}=f\left( r\right) dt~,\text{ \ }e^{1}=\frac{1}{f\left( r\right) }dr~,
\text{ \ }e^{m}=r\tilde{e}^{m}~,
\end{equation}
where $\tilde{e}^{m}$ denotes a vielbein for the base manifold $\sigma
_{\gamma }$. From the null torsion condition 
\begin{equation}
de^{a}+\omega _{\text{ \ }b}^{a}e^{b}=0~,
\end{equation}
we obtain the spin connection 
\begin{equation}
\omega ^{01}=f^{\prime }\left( r\right) e^{0},\text{ \ }\omega ^{m1}=f\left(
r\right) \tilde{e}^{m},\text{ \ }\omega ^{mn}=\tilde{\omega}^{mn}~,
\end{equation}
Now, by means of the change of coordinates $r=r_{+}\cosh \rho $, and using
the following representation of the gamma matrices 
\begin{equation}
\gamma ^{0}=i\sigma ^{2}\otimes \mathbf{1}~,\text{ \ }\gamma ^{1}=\sigma
^{1}\otimes \mathbf{1}~,\text{ \ }\gamma ^{m}=\sigma ^{3}\otimes \tilde{
\gamma}^{m}~,
\end{equation}
where $\sigma ^{i}$ are the Pauli matrices, and $\tilde{\gamma}^{m}$ are the
Dirac matrices in the base manifold $\sigma _{\gamma }$, along with the
following ansatz for the fermionic field 
\begin{equation}
\psi =e^{-i\omega t}\left( 
\begin{array}{c}
\psi _{1} \\ 
\psi _{2}
\end{array}
\right) \otimes \varsigma ~,
\end{equation}
where $\varsigma $ is a two component fermion. Also, doing the substitutions 
\cite{Das:1999pt} 
\begin{equation}\label{Das}
\psi _{1}\pm \psi _{2}=\sqrt{\frac{\cosh \rho \pm \sinh \rho }{\sinh \rho
\cosh ^{d-2}\rho }}\left( \psi _{1}^{\prime }\pm \psi _{2}^{\prime }\right)
~,
\end{equation}
and performing the change of variables $x=\tanh ^{2}\rho $, we obtain the
following equations 
\begin{eqnarray}\label{system}
2\left( 1-x\right) x^{1/2}\partial _{x}\psi _{1}^{\prime }+i\left( \frac{
\omega l}{\sqrt{\mu }}x^{-1/2}-\frac{\xi }{\sqrt{\mu }}x^{1/2}\right) \psi
_{1}^{\prime }+\left( i\left( \frac{\omega l}{\sqrt{\mu }}-\frac{\xi }{\sqrt{
\mu }}\right) +\frac{1}{2}+ml\right) \psi _{2}^{\prime } &=&0~,  \notag \\
2\left( 1-x\right) x^{1/2}\partial _{x}\psi _{2}^{\prime }-i\left( \frac{
\omega l}{\sqrt{\mu }}x^{-1/2}-\frac{\xi }{\sqrt{\mu }}x^{1/2}\right) \psi
_{2}^{\prime }+\left( -i\left( \frac{\omega l}{\sqrt{\mu }}-\frac{\xi }{
\sqrt{\mu }}\right) +\frac{1}{2}+ml\right) \psi _{1}^{\prime } &=&0~,
\end{eqnarray}
where $i\xi $ is the eigenvalue of the Dirac operator of the base manifold $
\sigma _{\gamma }$. By decoupling the system of equations and using 
\begin{equation}
\psi _{1}^{\prime }=x^{\alpha }\left( 1-x\right) ^{\beta }F\left( x\right)~, 
\end{equation}
with
\begin{equation}
\alpha =-\frac{i\omega l}{2\sqrt{\mu}}~, 
\end{equation}
\begin{equation}
\beta =-\frac{1}{2}\left( \frac{1}{2}+ml\right)~, 
\end{equation}
we obtain the following equation for $F\left(x\right)$
\begin{equation}
x\left( 1-x\right) F^{\prime \prime }\left( x\right) +\left( c-\left(
1+a+b\right) x\right) F^{\prime }\left( x\right) -abF\left( x\right) =0~, 
\end{equation}
whose solution is given by 
\begin{equation}
\psi _{1}^{\prime }=C_{1}x^{\alpha }\left( 1-x\right) ^{\beta }{_2F_1}\left(
a,b,c,x\right) +C_{2}x^{1/2-\alpha }\left( 1-x\right) ^{\beta }{_2F_1}\left(
a-c+1,b-c+1,2-c,x\right)~, 
\end{equation}
which has three regular singular point at $x=0$, $x=1$ and $x = \infty$. Here, $_2F_1(a,b,c;x)$ is a hypergeometric function and $C_1$, $C_2$ are constants and
\begin{equation}
a=\frac{1}{2}+\alpha +\beta -\frac{i\xi }{2\sqrt{\mu}}~, 
\end{equation}
\begin{equation}
b=\alpha +\beta +\frac{i\xi }{2\sqrt{\mu}}~, 
\end{equation}
\begin{equation}
c=\frac{1}{2}+2\alpha~. 
\end{equation}
Now, imposing boundary conditions at the horizon, i.e., that there is only ingoing modes, implies that $C_{2}=0$. Thus,  the solution can be written as
\begin{equation}
\psi _{1}^{\prime }=C_1x^{\alpha }\left( 1-x\right) ^{\beta }{_2F_1}\left(
a,b,c,x\right)~, 
\end{equation}
and by using the integrating factor 
$e^{-\int \frac{i}{2\left( 1-x\right) }\left( \frac{\omega l}{\sqrt{\mu }x}-
\frac{\xi }{\sqrt{\mu }}\right) dx}$, in Eq. (\ref{system}), we get the solution 
\begin{equation}
\psi _{2}^{\prime }=-\frac{C_1}{2}\left( -i\left( \frac{\omega l}{\sqrt{\mu }}-
\frac{\xi }{\sqrt{\mu }}\right) +\frac{1}{2}+ml\right) x^{-\alpha }\left(
1-x\right) ^{\alpha +\frac{i\xi }{2\sqrt{\mu }}}\int x^{\prime c-1}\left(
1-x^{\prime }\right) ^{a-c-1}{_2F_1}\left( a,b,c,x^{\prime }\right)
dx^{\prime }~.
\end{equation}
So, if we consider the relation
\begin{equation}
\int x^{c-1}\left( 1-x\right) ^{a-c-1}{_2F_1}\left( a,b,c,x\right) dx=\left(
1-x\right) ^{a-c}x^{c}\frac{{_2F_1}\left( a,b+1,c+1,x\right) }{c}~,
\end{equation}
and that  
\begin{equation}
-\frac{1}{2}\left( -i\left( \frac{\omega l}{\sqrt{\mu }}-\frac{\xi }{\sqrt{
\mu }}\right) +\frac{1}{2}+ml\right) =a-c~,
\end{equation}
$\psi _{2}^{\prime }$ can be rewritten as
\begin{equation}
\psi _{2}^{\prime }=\frac{a-c}{c}C_1x^{\alpha +1/2}\left( 1-x\right) ^{\beta
}{_2F_1}\left( a,b+1,c+1,x\right)~. 
\end{equation}

In this manner, the flux
\begin{equation}
\mathcal{F} =\sqrt{-g}\bar{\psi}\gamma ^{r}\psi~, 
\end{equation}
where, $\gamma ^{r}=e_{1}^{r}\gamma ^{1}$, $\bar{\psi}=\psi ^{\dagger
}\gamma ^{0}$,
$\sqrt{-g}=\left( l\sqrt{\mu}\right) ^{d-2}\left( 1-x\right) ^{-\left(
d-2\right) /2}$, 
and
$e_{1}^{r}\propto \frac{x^{1/2}}{\left( 1-x\right) ^{1/2}}$,
is given by 
\begin{equation}
\mathcal{F} \propto x^{1/2}\left( 1-x\right) ^{\left( -d+1\right) /2}\left(
\left\vert \psi _{1}\right\vert ^{2}-\left\vert \psi _{2}\right\vert
^{2}\right)~,
\end{equation}
and the flux as a function of  $\psi _{1}^{\prime }$ and $\psi _{2}^{\prime }$ is given by
\begin{equation}\label{flux}
\mathcal{F} \propto \left( \left\vert \psi _{1}^{\prime }\right\vert
^{2}-\left\vert \psi _{2}^{\prime }\right\vert ^{2}\right)~,
\end{equation}
where we have considered 
\begin{equation}
\left\vert \psi _{1}\right\vert ^{2}-\left\vert \psi _{2}\right\vert ^{2}=
\frac{\left( 1-x\right) ^{(d-1)/2}}{x^{1/2}}\left( \left\vert \psi
_{1}^{\prime }\right\vert ^{2}-\left\vert \psi _{2}^{\prime }\right\vert
^{2}\right)~,
\end{equation}
from (\ref{Das}), and doing $x=\tanh ^{2}\rho $. 

So, for $ml>-1/2$ and imposing
null flux at infinity $\rho \rightarrow \infty $, we obtain the following sets of
quasinormal frequencies 
\begin{eqnarray}
\omega &=&\frac{\xi }{l}-i\sqrt{\mu }\left( \frac{1}{2l}+\frac{2}{l}
n+m\right) ~,  \label{qnmf} \\
\omega &=&-\frac{\xi }{l}-i\sqrt{\mu }\left( \frac{3}{2l}+\frac{2}{l}
n+m\right) ~,  \notag
\end{eqnarray}
and, for $ml<-1/2$, we obtain 
\begin{eqnarray}
\omega &=&\frac{\xi}{l}-i\sqrt{\mu }\left( \frac{3}{2l}+\frac{2}{l}
n-m\right)~,  \label{frecuen} \\
\omega &=&-\frac{\xi}{l}-i\sqrt{\mu }\left( \frac{1}{2l}+\frac{2}{l}
n-m\right)~.  \notag
\end{eqnarray}
Where, we have considered the KummerÕs formula, \cite{M. Abramowitz}, 
\begin{eqnarray}
\nonumber {_2F_1}\left( a,b,c,x\right)  &=&\frac{\Gamma \left( c\right) \Gamma \left(
c-a-b\right) }{\Gamma \left( c-a\right) \Gamma \left( c-b\right) }
{_2F_1}\left( a,b,a+b-c,1-x\right) + \\
&&\left( 1-x\right) ^{c-a-b}\frac{\Gamma \left( c\right) \Gamma \left(
a+b-c\right) }{\Gamma \left( a\right) \Gamma \left( b\right) }{_2F_1}\left(
c-a,c-b,c-a-b+1,1-x\right) 
\end{eqnarray}
in Eq. (\ref{flux}). It is worth mentioning that these frequencies are similar to that of the BTZ black hole \cite
{Chan:1996yk, Cardoso:2001hn, Birmingham:2001pj} and that the imaginary part of the quasinormal frequencies is negative, which ensures
the stability of the black hole under fermionic perturbations. Also,  it is possible to observe that if $\mu =1$, we recover the QNMs of
the massless topological black holes in $d$-dimensions \cite
{LopezOrtega:2010uu}. Actually, if $\mu =1$ the metric (\ref
{dtopologicalmetric}) coincides with the metric of a massless topological
black hole. It is also interesting to observe that if $\mu \neq 1$ the QNMs (\ref{qnmf}) and (\ref{frecuen}) of fermionic perturbations for Chern-Simons
black holes have the imprint of the high curvature of the original theory. 
 According to the AdS/CFT correspondence the relaxation time $
\tau $ for a thermal state to reach thermal equilibrium in the boundary
conformal field theory is $\tau =1/\omega _{I}$ where $\omega _{I}$ is the
imaginary part of QNMs. As can be seen in relation to (\ref{qnmf}) and  (\ref{frecuen}), $\omega _{I}$
scales with $\sqrt{\mu }$. Then, depending on the sign of $\sigma $, $\mu $
can be larger of smaller than one. This means that the fermionic field will
decay faster or slower depending on the value of the curvature parameter $k$.

\section{Quasinormal modes of fermionic perturbations for some spacetimes
with torsion}
In this section we will study the quasinormal modes of fermionic
perturbations for some spacetimes with torsion. First, we will consider
the five-dimensional Chern-Simons black hole with torsion and then the BTZ black hole with torsion.

\subsection{Chern-Simons black hole with torsion}
The Chern-Simons black hole with torsion in five dimensions is a solution to
the Einstein-Gauss-Bonnet action \cite{Canfora:2007xs}, which, in terms of
differential forms is given by 
\begin{equation}
I=\int \epsilon _{abcde}\left( \frac{c_{0}}{5}e^{a}e^{b}e^{c}e^{d}e^{e}+
\frac{c_{1}}{3}R^{ab}e^{c}e^{d}e^{e}+c_{2}R^{ab}R^{cd}e^{e}\right) ~,
\end{equation}
where $e^{a}=e_{\mu }^{a}dx^{\mu }$ is the vielbein, and $R^{ab}=d\omega
^{ab}+\omega _{c}^{a}\omega ^{cb}$ is the curvature 2-form for the spin
connection $\omega ^{ab}=\omega _{\mu }^{ab}dx^{\mu }$. The Gauss-Bonnet
coupling is fixed as 
\begin{equation}
c_{2}=\frac{c_{1}^{2}}{4c_{0}}~.
\end{equation}
Thus, the theory possesses the maximum number of degrees of freedom and a
unique maximally symmetric vacuum \cite{Troncoso:1999pk, Crisostomo:2000bb},
and the Lagrangian can be written as a Chern-Simons form, \cite
{Chamseddine:1989nu}. The metric is given by 
\begin{equation}
ds^{2}=-f\left( r\right) ^{2}dt^{2}+\frac{dr^{2}}{f\left( r\right) ^{2}}
+r^{2}d\Omega ^{2}~,  \label{solution}
\end{equation}
with 
\begin{equation}
f\left( r\right) ^{2}=\frac{r^{2}}{l^{2}}-\mu ~,
\end{equation}
where $l=\sqrt{\frac{2c_{2}}{c_{1}}}$ is the AdS radius and $\Omega $ is a
3-dimensional base manifold, being this base manifold fixed but arbitrary. The metric (\ref{solution}) represents a black hole for $\mu >0$ and the torsion
2-form is given by the expression 
\begin{equation}
T^{m}=-\frac{\delta }{r}\epsilon ^{mnp}e_{n}e_{p}~,
\end{equation}
where $\delta ~$is an integration constant. Here, the indices $m,n,...~\ $
refers to $\Omega $. Therefore, the non-vanishing components of the torsion
have support along the base manifold only. The spin connection results in
\begin{equation}
\omega ^{01}=f^{\prime }\left( r\right) e^{0},\text{ \ }\omega ^{n1}=f\left(
r\right) \tilde{e}^{n},\text{ \ }\omega ^{mn}=\tilde{\omega}^{mn}+K^{mn}~,
\end{equation}
where the contorsion reads 
\begin{equation}
K^{mn}=\frac{\delta }{r}\epsilon ^{mnp}e_{p}~.  \label{contorsion}
\end{equation}

In this case, a minimally coupled fermionic field to curvature in
the background of the five-dimensional Chern-Simons black hole with torsion is
given by the Dirac equation with torsion, \cite{Adak:2011md}
\begin{equation}  \label{TD}
^{\star }\gamma \wedge \left( \nabla-\frac{1}{2}K_{\text{ \ \ \ }
a}^{ab}e_{b}\right) \psi +m\psi ^{\star }1=0~.
\end{equation}
However, for the solution (\ref{contorsion}) $K_{\text{ \ \ \ }a}^{ab}=0$,
the Dirac equation (in a coordinate base \{$dx^{\mu }$\}) simplifies to 
\begin{equation}
\left( \gamma ^{\mu }\nabla _{\mu }+m\right) \psi =0~,
\end{equation}
where the effect of the torsion is included in the covariant derivative
through the spin connection. As in section III, we consider the change of
coordinates $r=r_{+}\cosh \rho $ and we use the following ansatz for the
fermionic field 
\begin{equation}
\psi =e^{-i\omega t}\left( 
\begin{array}{c}
\psi _{1} \\ 
\psi _{2}
\end{array}
\right) \otimes \varsigma ~.
\end{equation}
The substitutions (\ref{Das}), with $d=5$ and  
the following representation for the Dirac matrices 
\begin{equation}
\gamma ^{0}=i\sigma ^{2}\otimes \mathbf{1},\text{ \ }\gamma ^{1}=\sigma
^{1}\otimes \mathbf{1},\text{ \ }\gamma ^{m}=\sigma ^{3}\otimes \tilde{\gamma
}^{m}~,
\end{equation}
permit us to obtain the 
equations 
\begin{eqnarray}
2\left( 1-x\right) x^{1/2}\partial _{x}\psi _{1}^{\prime }+i\left( \frac{
\omega l}{\sqrt{\mu }}x^{-1/2}-\frac{\left( \xi +\frac{3}{2}\delta \right) }{
\sqrt{\mu }}x^{1/2}\right) \psi _{1}^{\prime }+\left( i\left( \frac{\omega l
}{\sqrt{\mu }}-\frac{\left( \xi +\frac{3}{2}\delta \right) }{\sqrt{\mu }}
\right) +\frac{1}{2}+ml\right) \psi _{2}^{\prime } &=&0~,  \notag \\
2\left( 1-x\right) x^{1/2}\partial _{x}\psi _{2}^{\prime }-i\left( \frac{
\omega l}{\sqrt{\mu }}x^{-1/2}-\frac{\left( \xi +\frac{3}{2}\delta \right) }{
\sqrt{\mu }}x^{1/2}\right) \psi _{2}^{\prime }+\left( -i\left( \frac{\omega l
}{\sqrt{\mu }}-\frac{\left( \xi +\frac{3}{2}\delta \right) }{\sqrt{\mu }}
\right) +\frac{1}{2}+ml\right) \psi _{1}^{\prime } &=&0~,
\end{eqnarray}
where $i\xi $ is the eigenvalue of the Dirac operator of the base manifold $
\Omega $, and we have used $x=\tanh ^{2}\rho $. So, imposing null flux at infinity, for $ml>-1/2$  we
get the quasinormal frequencies 
\begin{eqnarray}
\omega &=&\frac{\left( \xi +\frac{3}{2}\delta \right) }{l}-i\sqrt{\mu }
\left( \frac{1}{2l}+\frac{2}{l}n+m\right) , \\
\omega &=&-\frac{\left( \xi +\frac{3}{2}\delta \right) }{l}-i\sqrt{\mu }
\left( \frac{3}{2l}+\frac{2}{l}n+m\right) ~, \notag
\end{eqnarray}
and for $ml<-1/2$ we obtain frequencies similar to (\ref{frecuen}) making the
change $\xi \rightarrow \xi +\frac{3}{2}\delta $. Therefore, we see that the
effect of torsion, represented by the term $\delta $, is  modifies the real
part of the quasinormal frequencies, which modify the oscillation frequency
of the field. Also, this spacetime is stable because the imaginary part
is negative and turns out  to be the same as the Chern-Simons black hole without
torsion. Note that, depending on the value of $\delta $, a value
of $\xi $, $\xi =-3/2\delta $ may exist, for which the quasinormal frequencies are
purely imaginary, and therefore the field decay in time without oscillation.

\subsection{\protect\bigskip BTZ black hole with torsion}
Now, we consider the BTZ black hole with torsion \cite{Garcia:2003nm}, which is
a solution of the topological Mielke-Baekler model for 3D gravity \cite
{Mielke:1991nn} 
\begin{equation}
I=\int 2ae^{a}R_{a}-\frac{\Lambda }{3}\epsilon _{abc}e^{a}e^{b}e^{c}+\alpha
_{3}\left( \omega ^{a}d\omega _{a}+\frac{1}{3}\epsilon _{abc}\omega
^{a}\omega ^{b}\omega ^{c}\right) +\alpha _{4}e^{a}T_{a}~,
\end{equation}
where, the first term corresponds to the usual Einstein-Cartan action with $a=
\frac{1}{16\pi G}$, the second term is a cosmological term, the following
term is the Chern-Simons action for the Lorentz connection, and the last term
is a torsion counterpart of the first one. Also, the dual expressions $
\omega _{a}$ and $R_{a}$ are defined by $\omega ^{ab}=\epsilon ^{abc}\omega
_{c}$ and $R^{ab}=\epsilon ^{abc}R_{c}$, respectively. We have used the
notation of \cite{Cvetkovic:2007fi} with $\eta _{ab}=diag\left( -,+,+\right) 
$, and in the next we will consider $8G=1$. The BTZ black hole with torsion is
described by the metric 
\begin{equation}
ds^{2}=-N\left( r\right) ^{2}dt^{2}+\frac{1}{N\left( r\right) ^{2}}
dr^{2}+r^{2}\left( d\phi +N_{\phi }dt\right) ^{2}~,
\end{equation}
where 
\begin{equation}
N\left( r\right) ^{2}=\frac{r^{2}}{l^{2}}-M+\frac{J^{2}}{4r^{2}},\text{
\ }N_{\phi }\left( r\right) =\frac{J}{2r^{2}}~.
\end{equation}
and $\Lambda _{eff}=-\frac{1}{l^{2}}$, with $\Lambda _{eff}=q-\frac{1}{4}
p^{2}$, where we have $q=-\frac{\alpha _{4}^{2}+a\Lambda }{\alpha _{3}\alpha
_{4}-a^{2}}$ and $p=\frac{\alpha _{3}\Lambda +\alpha _{4}a}{\alpha
_{3}\alpha _{4}-a^{2}}$ are constants. For simplicity, we will analyze the
case with $J=0$. In this case, the event horizon is located at $r_{+}=l\sqrt{
M}$. Now, considering the diagonal vielbein 
\begin{equation}
e^{0}=N\left( r\right) dt~,\text{ \ }e^{1}=\frac{1}{N\left( r\right) }dr~,
\text{ \ }e^{2}=rd\phi ~,
\end{equation}
the spin connection reads 
\begin{equation}
\omega ^{01}=\frac{r}{l^{2}}dt-\frac{p}{2}rd\phi ,\text{ \ }\omega ^{02}=
\frac{p}{2N\left( r\right) }dr,\text{ \ }\omega ^{12}=-N\left( r\right)
d\phi +\frac{pN\left( r\right) }{2}dt~.
\end{equation}
For $p=0$, the torsion vanishes and the spin connection becomes the
Levi-Civita spin connection. In this case, the Dirac equation with torsion, by considering the change of
coordinates $r=r_{+}\cosh \rho $, and the following representation of
the gamma matrices 
\begin{equation}
\gamma ^{0}=i\sigma ^{2},\text{ \ }\gamma ^{1}=\sigma ^{1},\text{ \ }\gamma
^{2}=\sigma ^{3}~,
\end{equation}
where $\sigma ^{i}$ are the Pauli matrices, along with the following ansatz
for the fermionic field
\begin{equation}
\psi =e^{-i\omega t}e^{i\xi \phi }\left( 
\begin{array}{c}
\psi _{1} \\ 
\psi _{2}
\end{array}
\right) ~,
\end{equation}
and Eq. (\ref{Das}) for $d=3$ and    
$x=\tanh ^{2}\rho $, permit us to write
the following equations 
\begin{eqnarray}
2\left( 1-x\right) x^{1/2}\partial _{x}\psi _{1}^{\prime }+i\left( \frac{
\omega l}{\sqrt{M}}x^{-1/2}-\frac{\xi }{\sqrt{M}}x^{1/2}\right) \psi
_{1}^{\prime }+\left( i\left( \frac{\omega l}{\sqrt{M}}-\frac{\xi }{\sqrt{M}}
\right) +\frac{1}{2}+ml+\frac{3pl}{4}\right) \psi _{2}^{\prime } &=&0~, 
\notag \\
2\left( 1-x\right) x^{1/2}\partial _{x}\psi _{2}^{\prime }-i\left( \frac{
\omega l}{\sqrt{M}}x^{-1/2}-\frac{\xi }{\sqrt{M}}x^{1/2}\right) \psi
_{2}^{\prime }+\left( -i\left( \frac{\omega l}{\sqrt{M}}-\frac{\xi }{\sqrt{M}
}\right) +\frac{1}{2}+ml+\frac{3pl}{4}\right) \psi _{1}^{\prime } &=&0~.
\end{eqnarray}
The difference in these equations with respect to the torsionless BTZ black
hole is the term $\frac{3pl}{4}$. This term redefines the mass of the
fermionic field and therefore modifies the quasinormal frequencies of the
field. The quasinormal frequencies for $ml+\frac{3pl}{4}>-1/2$ (imposing null
flux at infinity $\rho \rightarrow \infty $) are given by 
\begin{eqnarray}
\omega  &=&\frac{\xi }{l}-i\sqrt{M}\left( \frac{1}{2l}+\frac{2}{l}n+m+\frac{
3p}{4}\right) ~, \\
\omega  &=&-\frac{\xi }{l}-i\sqrt{M}\left( \frac{3}{2l}+\frac{2}{l}n+m+\frac{
3p}{4}\right) ~,  \notag
\end{eqnarray}
and for $ml+\frac{3pl}{4}<-1/2$ frequencies similar to (\ref{frecuen}) are
obtained by making the changes $ml\rightarrow ml+\frac{3pl}{4}$ and $\mu
\rightarrow M$. The effect of the torsion is to modify the imaginary part of
these frequencies in contrast with the five-dimensional Chern-Simons case,
where the torsion effect was to modify the real part of the quasinormal
frequencies. This modification that happens for fermionic fields is absent
for scalar fields, where there is no contribution by the torsion to the
quasinormal modes. Note that if $\xi =0$, the quasinormal frequencies are
purely imaginary, and therefore the field decay in time without oscillation \footnote{Recently, in Ref. \cite{Ma:2013eaa} the QNFs obtained in this article for the BTZ black hole with torsion were used to compute the entropy spectrum of this black hole, and was shown that it is equally spaced.}.

\section{Conclusions}
In this work, we have calculated the QNMs of fermionic perturbations for
Chern-Simons black holes with spherical, hyperbolic and plane topology.
Here, we have considered as boundary conditions (to determine the QNMs) that
there are only ingoing modes at the horizon. However, it is known that at
infinity their depend on the asymptotic behavior of spacetime. For asymptotically AdS spacetimes  the potential diverges and in this manner it
is required for the field be null, at infinity. However, by establishing
these Dirichlet boundary conditions, not
all QNMs are obtained. It is known that the QNMs of the BTZ black hole under Dirichlet boundary conditions permit to obtain only  a set of QNMs, for positive
masses of the scalar field. However,  there is another set of QNMs for a range of
imaginary masses which are allowed because the propagation of the scalar field
is stable, according to the Breitenlohner-Freedman limit, \cite{Breitenlohner:1982bm,Breitenlohner:1982jf}.
This set of QNMs, just as the former, can be obtained by requesting the flux
to vanish at infinity, which are known as Neumann boundary conditions. Remarkably, for fermionic perturbations there is no Breitenlohner-Freedman
limit. However, it is possible to consider Neumann boundary conditions
because Dirichlet boundary conditions would lead to the absence of QNMs for a
range of masses, without a physical reason for this absence, \cite{Birmingham:2001pj}. Here, we have considered Neumann boundary conditions at
infinity for Chern-Simons black holes and we have found that the Dirac QNMs depend
on the highest power of curvature present in the Lanczos-Lovelock theories,
such as the quasinormal modes of scalar perturbations. Then, we have found
the quasinormal modes of fermionic perturbations for the five-dimensional
Chern-Simons black hole with torsion and for the BTZ black hole with torsion
analytically.

We have found that these black holes are stable under fermionic perturbations, due to the imaginary part of the Dirac QNFs is negative. Also, we have found that the effect of the torsion on the Dirac QNFs of five-dimensional Chern-Simons black hole is to modify their real part (oscillatory frequency), in comparison with the Dirac QNFs of five-dimensional Chern-Simons black hole without torsion. Therefore, the relaxation time of the perturbation doesn't depend on the torsion. On the other hand, the effect of the torsion on the Dirac QNFs of  BTZ black hole with torsion is to modify the imaginary part (damping). Therefore, in this case, the relaxation time for the
decay of the black hole perturbation depend on the torsion value. It is worth mentioning that both metrics are very similar. However,  the torsion for the Chern-Simons black hole only exist in the three-dimensional base manifold, which is endowed with a fully antisymmetric torsion. In contrast with the BTZ black hole with torsion, where the torsion exist in the full spacetime. In this case, the torsion cannot be restricted to the base manifold. So, the fact that in the Chern-Simons case the torsion considered is completely antisymmetric and also is restricted to live in the three-dimensional base manifold is the reason why the torsion only enters into the real part of the QNFs. In this manner, in order to determine if the torsion could affect the imaginary part of the QNFs too, as it happens in the BTZ black hole with torsion, would be interesting to study Chern-Simons black holes solutions endowed with a more general torsion.

In the context of AdS/CFT correspondence, we can appreciate that the relaxation time for a thermal state to reach thermal equilibrium in the boundary
conformal field theory is modified by the torsion in three dimensions. Which can be established by assuming that the QNFs of the black hole are related with the poles of the retarded correlation function of the corresponding perturbations of the dual conformal field theory as in AdS spacetime without torsion \cite{Birmingham:2001pj}, which is matter of a current investigation.

\section*{Acknowledgments}
This work was funded by Comisi{\'o}n Nacional de  Investigaci{\'o}n Cient{\'i}fica y Tecnol{\'o}gica through FONDECYT Grant 11121148 (Y.V.). P.G. wish to thank the kind hospitality at the Departamento de F{\'i}sica, Universidad de La Serena. The authors acknowledge the referee for useful suggestions in order to improve this paper.

\end{document}